# (Information) Paradox Lost

Tim Maudlin

Department of Philosophy
New York University
New York, NY 10003

**Abstract**

Since Stephen Hawking's original 1975 paper on black hole evaporation there has been a consensus that the problem of "loss of information" is both deep and troubling, and may hold some conceptual keys to the unification of gravity with the other forces. I argue that this consensus view is mistaken. The so-called "information loss paradox" arises rather from the inaccurate application of foundational principles, involving both mathematical and conceptual errors. The resources for resolving the "paradox" are familiar and uncontroversial, and have been pointed out in the literature. The problem ought to have been dismissed 40 years ago. Recent radical attempts to "solve" the problem are blind alleys, solutions in search of a problem.

Introductory Incitement

There is no "information loss" paradox. There never has been. If that seems like a provocation, it's because it is one. Few problems have gotten as much attention in theoretical foundations of physics over the last 40 years as the so-called information loss paradox. But the "solution" to the paradox does not require any new physics that was unavailable in 1975, when Stephen Hawking posed the problem [1]. Indeed, the complete solution can be stated in a single sentence. And it has been pointed out, most forcefully by Robert Wald.

The claim above ought to be frankly incredible. How could a simple solution have gone unappreciated by so many theoretical physicists—among them very great physicists—over such a long period of time? Probably no completely satisfactory non-sociological explanation is possible. But in order to help clarify the situation it is useful to point out two potential confusions, one mathematical and the other conceptual. Each of these misunderstandings is easy to grasp, and when they are taken account of it is the problem itself that evaporates.

What, then, should this paper provoke? One of two things: either talk of an "information loss paradox" ought to cease, or some error in what follows ought to be articulated clearly and precisely. If the former, then along with the information loss paradox talk of "black hole complementarity", "firewalls", and "EPR = ER" ought to stop as well. If the latter, then the exact nature of the paradox will have been made clearer, and further discussion can be more focused and productive.

What is the paradox supposed to be?

Despite the usual moniker, the paradox is only tangentially about information, and the analytical tools of information theory (Shannon or otherwise) are not relevant to it. A much better characterization of the problem adverts to an apparent breakdown of either determinism or of unitary evolution of the quantum state. To be more precise on the first point, there is an apparent breakdown of retrodictability: from the quantum state of the universe "after a black hole has evaporated" one cannot recover the state "before the black hole evaporated", including the time while the black hole was first forming. This is often stated by saying "after the black hole evaporates" one cannot determine what sort of stuff originally came together to form it[1]. If, for example, a book was thrown into the event horizon "before the evaporation" one cannot determine the contents of the book from the physical state "after the evaporation". (The point of the liberal use of scare quotes here and in what follows will become clear anon.)

In an obvious sense, this failure of retrodictability entails a loss of information. The physical state "before the evaporation" contains information about what fell in, while the physical state "after the evaporation" does not. If the earlier state were retrodictable from the later one then the earlier one, with all of its information, could in principle be recovered from the later one. This also entails a failure of determinism in the backward time direction: different possible earlier states must be compatible with the same later state and the laws of physics. Hence the laws do not fix a unique transition from later to earlier: determinism fails in the backward direction of time.

---

[1] Here and elsewhere I characterize certain claims as "what is often said" or "what is typically said" or "what is usually said" without providing citations. These claims are so pervasive that a complete enumeration would be impossible, and a partial one unenlightening. Just check around in the literature.

What about the supposed breakdown of unitarity? Sometimes the paradox is presented by arguing that the later state is a mixed state while the earlier one is a pure state. But such pure-to-mixed transitions violate the unitarity of the dynamical law. Since the fundamental dynamics of quantum theory is unitary, it cannot yield a pure-to-mixed transition, and so cannot deal with black hole evaporation.

If one interprets the late mixed state as a proper statistical mixture then we also again get a failure of determinism, now in the forward time direction. That is, if we regard the mixed state as representing incomplete information about the outcome of a stochastic evolution, with different possible outcomes weighted by their probabilities, then we accept that the same early state can give rise to different possible later states. Determinism fails again.

These are the basic principles that are invoked in arguing for a paradox. Quantum theory is supposed to provide a deterministic, unitary, retrodictable evolution of the quantum state, but the evolution in the presence of an evaporating black hole appears to be indeterministic, non-unitary, and non-retrodictable. We have seen how information about what fell into the black hole is lost in the sense that the earlier state cannot be recovered by analysis from the later state.

There is an obvious sense in which any dynamical evolution that is deterministic in both time directions "preserves information". In such a case, the value of the state at any time implies the value of the state at any other time. From the complete physical state at any time everything that ever happens anywhere can in principle be recovered. This is the point that Laplace made about deterministic physics:

> We may regard the present state of the universe as the effect of its past and the cause of its future. An intellect which at a certain moment would know all forces that set nature in motion, and all positions of all items of which nature is composed, if this intellect were also vast enough to submit these data to analysis, it would embrace in a single formula the movements of the greatest bodies of the universe and those of the tiniest atom; for such an intellect nothing would be uncertain and the future just like the past would be present before its eyes [2] p. 4.[2]

---

[2] Although Laplace is usually given credit for this thought, it was even more clearly and precisely articulated earlier by Roger Boscovich. "Now, if the law of forces were known, & the position, velocity & direction of all points at any given instant, it would be possible for a mind of this type to foresee all the necessary subsequent motions & states, & to predict all the phenomena that necessarily followed from them. It would

No concepts from information theory are required to state this principle. The real issue concerns determinism forward in time and retrodictability backward in time.

## Determinism, Retrodictability and Unitarity

The information loss paradox is set in a relativistic space-time. Under what conditions should one have any right to expect that a dynamical transition will be unitary, deterministic, or retrodictable in that setting?

The answer is sharp and has been known for a long time. The only transitions for which these features can be expected are transitions (either backward or forward in time) from the state on one Cauchy surface in the space-time to another. Even if the theory is fully deterministic in Laplace's (Boscovich's) sense, states defined on anything less than a Cauchy surface may fail to have these properties.

Recall: A Cauchy surface is a locus of events that every inextensible timelike curve intersects exactly once. As Wald ([4] p. 181) points out, even in a fully deterministic setting, the transition from the state on a Cauchy surface to the state on a non-Cauchy surface may be deterministic forward in time but not retrodictable. In Minkowski space-time, for example, let $\Sigma_1$ in Figure 1 be the locus of events with $t = 0$ in some set of Lorentz coordinates and $\Sigma_2$ be the hyperboloid $(t-1)^2 - x^2 - y^2 - z^2 = 1$ with $t > 0$. Suppose there is only one light ray in the space-time, which passes through the origin. Since $\Sigma_2$ lies entirely inside the future lightcone of the event (1,0,0,0) and that light cone in turn lies entirely inside the future light cone of the origin, the light ray will never intersect $\Sigma_2$. Although the physical state on $\Sigma_1$ "contains the information" that the light ray went through the origin, and although the physical state on $\Sigma_2$ is completely determined by the state on $\Sigma_1$, the existence and trajectory of the light ray never registers on $\Sigma_2$. Even if you know everything about the state of $\Sigma_2$ you could not retrodict from it the trajectory, or even existence, of the light ray.

---

be possible from a single arc described by any point in an interval of continuous time, no matter how small, which was sufficient for a mind to grasp, to determine the whole of the remainder of such a continuous curve, continued to infinity on either side". See [3].

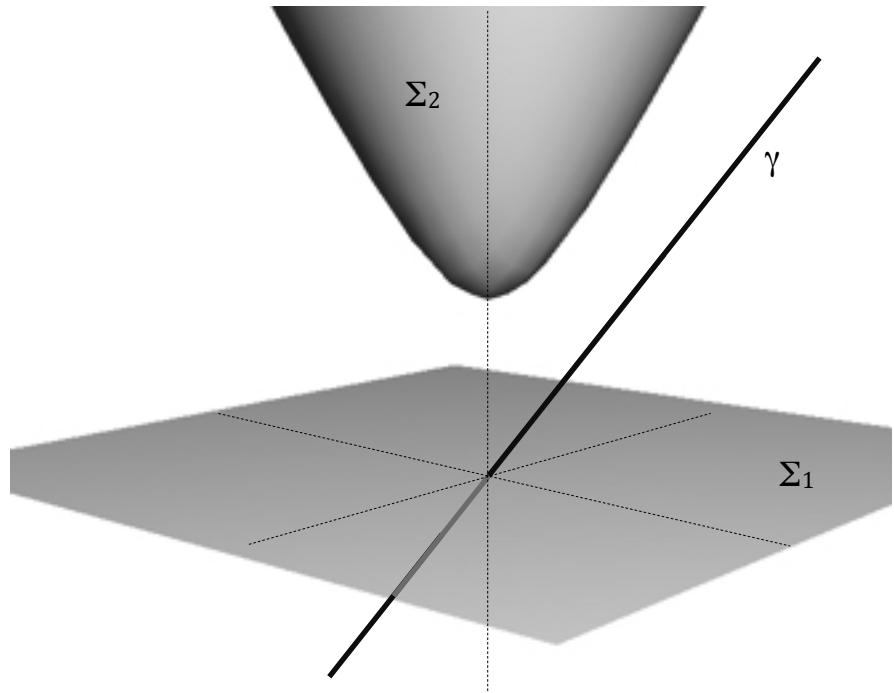

Figure 1: Two inextensible spacelike surfaces in Minkowski space-time

Why does retrodiction fail here? Because $\Sigma_2$ isn't a Cauchy surface. Some timelike curves, such as the solutions to $x^2 - t^2 = 1$, fail to intersect $\Sigma_2$. But electrodynamics is a completely deterministic theory (in both time directions) in Minkowski space-time. The moral is that the only transitions one can rightly expect to be deterministic, or to be retrodictable, or to "preserve information", are transitions from the complete physical state on one Cauchy slice to the complete state on another Cauchy slice. This is true already even in Special Relativity independently of any features of quantum theory or of gravity.

To repeat: in a relativistic theory, the *only* dynamical transitions one has any right to expect to be deterministic and retrodictable and unitary are transitions from the state on one Cauchy surface to the state on another. A state on $\Sigma_1$ can deterministically give rise to a state on $\Sigma_2$ from which the state on $\Sigma_1$ cannot be recovered; a pure state on $\Sigma_1$ can give rise to an (improper) mixed state on $\Sigma_2$ if the state on $\Sigma_2$ is constructed by tracing out degrees of freedom that are not recorded on it.

A Cauchy surface is what corresponds in Relativity to "a certain moment" for Laplace, or "a given instant" for Boscovich, or a "moment of duration" for Newton. Newton's explication is the most arresting:

> For we do not ascribe various durations to the different parts of space, but say that all endure together. The moment of duration is that same at Rome and at London, on the Earth and on the stars, and throughout all the heavens. And just as we understand any moment of duration to be diffused throughout all spaces….([5] p. 113)

A Cauchy slice too is "diffused throughout all spaces" as the diagram indicates.

Non-Evaporating Black Holes

Black hole solutions to the Einstein Field Equations have been recognized since Karl Schwartzschild found his solution just months after the publication of the theory in 1915. Yet there was no issue about "information loss" in a black hole until the postulation of evaporating black holes by Hawking [1]. Evidently whatever the information loss paradox is supposed to be, it does not arise for a non-evaporating black hole.

In the non-evaporating case, "information" can indeed pass from outside the event horizon to inside the event horizon and so effectively be "lost" to observers who stay outside. But that presents nothing paradoxical. As Hawking writes:

> This loss of information wasn't a problem in the classical theory. A classical black hole would last forever and the information could be thought of as preserved inside it, but just not very accessible. However, the situation changed when I discovered that quantum effects would cause a black hole to radiate at a steady rate. [6].

Let's make Hawking's observation precise.

Figure 2 is the Penrose diagram of a non-evaporating black hole. The shaded region represents the matter that collects together to form the black hole. Two Cauchy surfaces are indicated: $\Sigma_1$, which does not enter the event horizon and $\Sigma_2$, which does. Since these are Cauchy surfaces, we expect the dynamical transition from the state on one to the state on the other to be deterministic, retrodictable, and unitary. And it is. So $\Sigma_2$ must "contain information about what fell into the event horizon". Further, insofar as we can localize "information" at all (about which later), this information is located on the part of the Cauchy Surface that lies inside the horizon.

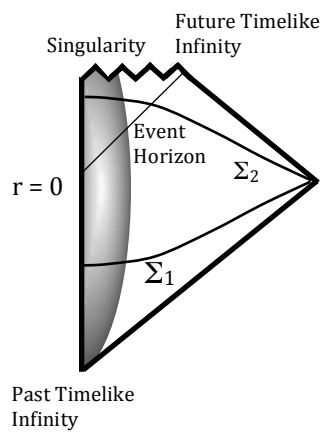

Figure 2: Penrose Diagram of a Non-Evaporating Black Hole

It is clear that both $\Sigma_1$ and $\Sigma_2$ are Cauchy surfaces. Every inextensible timelike curve "originates" at past timelike or null infinity and "terminates" either at the singularity or at future timelike or null infinity. Every such curve intersects both $\Sigma_1$ and $\Sigma_2$, and does so exactly once.

There are scare quotes around "originates" and "terminates" because neither of these points on the diagram represents a real physical event. The singularity line represents an open edge of the manifold, as do the points "at infinity". It will be useful to keep in mind that not every point in a Penrose diagram represents an event (or collection of events) in space-time.

Hawking's description of the situation is a bit too mild. It is not that the information about what fell in *could be* regarded as persisting inside the event horizon, but rather it *must be* so regarded. $\Sigma_2$ must have a part in the interior of the event horizon in order to be a Cauchy surface. It is on that part that the infalling matter registers. If a book is thrown into the black hole and one asks where the information in the book resides on $\Sigma_2$, the answer is clear: it resides where the book intersects $\Sigma_2$. So information is never lost and the states on Cauchy surfaces always evolve into each other deterministically, retrodictably, and unitarily. No problem.

## Evaporating Black Holes

As Hawking goes on to relate, the information loss problem only arises with the postulation of evaporating black holes.

> However, the situation changed when I discovered that quantum effects would cause a black hole to radiate at a steady rate. At least in the approximation I was using the radiation from the black hole would be completely thermal and would carry no information. So what would happen to all that information locked inside a black hole that evaporated away and disappeared completely? It seemed the only way the information could come out would be if the radiation was not exactly thermal but had subtle correlations. No one has found a mechanism to produce correlations but most physicists believe one must exist. If information were lost in black holes, pure quantum states would decay into mixed states and quantum gravity would not be unitary. [6]

The cogency of Hawking's 1975 argument that the black hole should lose mass and eventually evaporate is not completely evident, but that is a matter for another time. For the remainder of this paper we will accept that the existence of Hawking radiation implies evaporation.

Hawking's 1975 paper contains the first Penrose diagram of an evaporating black hole (Figure 3).

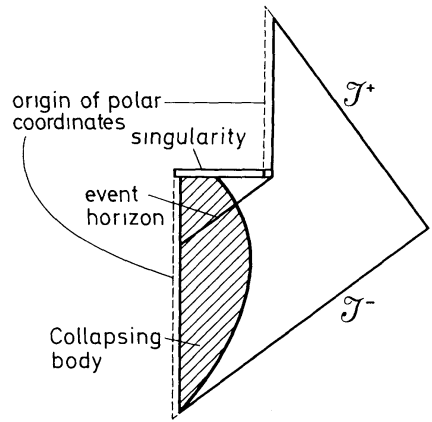

The Penrose diagram for a gravitational collapse followed by the slow evaporation and eventual disappearance of the black hole, leaving empty space with no singularity at the origin

Figure 3: Hawking's Original Penrose Diagram, with his Caption

There are a couple of notable features about this diagram. One is that it is vague about the status of the corners where r = 0 and the singularity intersect. r = 0 everywhere else represents actual events in space-time. The line labeled "singularity" in the diagram does not. Those points on the diagram do not represent actual physical locations in space-time: there is no there there. What about the two corners?

The left corner clearly does not represent an actual physical event. We should think of the r = 0 axis as having an open edge in that part of the diagram. This is clear because the location of r = 0 is a matter of convention: in different coordinate systems, r = 0 indicates different events.

The right corner is much more interesting. We take it that that corner in the diagram does represent a physical event, a point in space-time. We will call this the Evaporation Event. The geometry of the Evaporation Event comes in for more discussion below. If that point on the diagram failed to represent a physical event, then the light on the event horizon would end at the singularity, and the singularity would be naked there.

As with the non-evaporating black hole diagram, the top and bottom corners labeled "timelike infinity" and the edges labeled $\mathscr{I}^+$ and $\mathscr{I}^-$ do not represent points in space-time. But every inextensible timelike curve will originate at either Past Timlike Infinity or $\mathscr{I}^-$, and every such curve will terminate either at the singularity, at Future Timelike Infinity, or at $\mathscr{I}^+$. All inextensible timelike curves are topologically open in both directions.

Taking these observations into account, we can draw a more explicit Penrose diagram as in figure 4:

Figure 4: A More Explicit Penrose Diagram

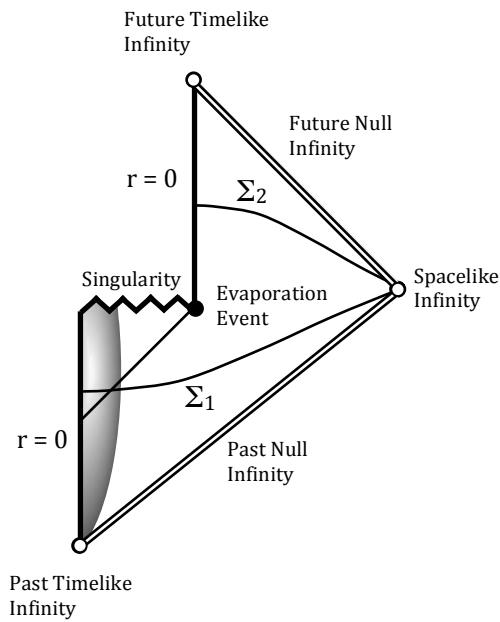

The solid black lines and solid black circle in the diagram represent actual space-time points, while the open circles and striped lines do not correspond to space-time points and neither, of course, does the wavy singularity line. We have also added the edgeless space-like surfaces $\Sigma_1$ and $\Sigma_2$. The points in the interior of the Penrose diagram all represent spacelike topological 2-spheres in the physical space-time. (I have not used the more technical names $i^-$, $i^+$, $i^0$, $\mathscr{I}^-$ and $\mathscr{I}^+$ since those only properly belong to diagrams of asymptotically flat space-times, and this space-time is not asymptotically flat.)

When Hawking writes in the caption to his diagram that the evaporation leaves "empty space with no singularity at the origin" he is clearly referring to a slice like $\Sigma_2$. Of course, $\Sigma_2$ is not "empty": it is supposed to contain the Hawking radiation that carries away energy from the black hole. Hawking's argument for information loss is then simple: if the Hawking radiation is thermal, with a spectrum depending only on the mass, charge and angular momentum of the black hole, then it carries no information about exactly what sort of matter formed the black hole save for its mass, charge and angular momentum. Information present on $\Sigma_1$ is absent on $\Sigma_2$. Information is lost.

Now we can make Wald's key observation:

$\Sigma_2$ is not a Cauchy surface.

This point cannot be overemphasized. It has been noticed in the literature, but the significance of it has evidently not been widely appreciated.

It is obvious—trivial—that $\Sigma_2$ is not a Cauchy surface. Any inextensible timelike curve that originates at past timelike infinity or $\mathscr{I}^-$ and terminates at the singularity fails to intersect $\Sigma_2$. $\Sigma_1$, in contrast, is a Cauchy surface. Since every inextensible timelike curve "originates" at past timelike infinity or $\mathscr{I}^-$, every one intersects $\Sigma_1$ exactly once. So the transition from the physical state on $\Sigma_1$ to the physical state on $\Sigma_2$ is a Cauchy-to-non-Cauchy transition. *That means that there never has been any grounds to expect the transition to be either retrodictable or unitary.* Quantum theory does not imply, and never has, that such a transition must be retrodictable or unitary or "preserve information". There were no grounds in 1975, and remain no grounds today, to expect information to be preserved. *There was never any information loss paradox based in fundamental properties of quantum theory and relativity.*

This paper could end here. Nothing that I have pointed out is in the least controversial, but what follows from it is that *forty years of effort has been directed at a non-problem*. And looking for the solution of a non-problem is likely to be futile. Black hole complementarity, firewalls, EPR = ER are all solutions in search of a problem. There is no reason to think there is any such problem to solve.

What we have left to us, then, are two important questions. First, if the information about what originally formed the black hole (and more generally whatever passed through the event horizon) is not present on $\Sigma_2$, where is it? And second, how did so many prominent and brilliant physicists manage to get so confused? The answers to these questions may be linked.

The Two Theorems

Once we realize that determinism, retrodictability and unitarity are only to be expected for Cauchy-to-Cauchy transitions, we understand why the "missing information" need not be on $\Sigma_2$, since it is not a Cauchy surface. So what's the situation with respect to Cauchy surfaces in the space-time of Figure 4?

Two theorems proven by Robert Geroch [7] bear on this issue. The first theorem proves that any space-time that admits of a Cauchy surface admits of a foliation into Cauchy surfaces. Since the total information about the physical world is only preserved on Cauchy-to-Cauchy transitions, our first question is what such a foliation of our evaporating black hole space-time looks like.

The space-time obviously admits of a Cauchy surface since $\Sigma_1$ is one. Let's start with $\Sigma_1$, then, and proceed from there. Foliating backwards to past timelike infinity it trivial: just keep slicing more-or-less parallel to $\Sigma_1$. But what happens as we try to continue the foliation forward in time?

Things go smoothly until a leaf of the foliation hits the Evaporation Event. Let's call that leaf $\Sigma_{critical}$. Then what? We have no choice but to continue by using *disconnected* Cauchy surfaces. Our friend $\Sigma_2$ is one part of such a Cauchy surface, and will catch all of the inextensible timelike curves that run from Past Timelike or Null Infinity to Future Timelike or Null Infinity, i.e. all of the curves that never pass the event horizon. In order to construct a complete Cauchy surface we need to add to $\Sigma_2$ another piece that will catch all of the inextensible timelike curves that run from Past Timelike or Null Infinity to the singularity. That is, we need to append a piece that lives *inside the event horizon*. Consider a spacelike line that goes from a point on r = 0 to the Evaporation Event. Let it be open at the Evaporation Event. In effect, the Evaporation Event functions for all the slices inside the event horizon in exactly the same way as Spacelike Infinity functions for the spacelike slices outside the horizon, with the difference that the Evaporation Event represents a real, physical event in the space-time. Such a line will catch all of the inextensible timelike curves that enter the event horizon and go to the singularity. Let's call one such line "$\Sigma_{2in}$", and rename our original $\Sigma_2$ "$\Sigma_{2out}$". Then $\Sigma_{2in} \cup \Sigma_{2out}$ forms a Cauchy surface. See Figure 5.

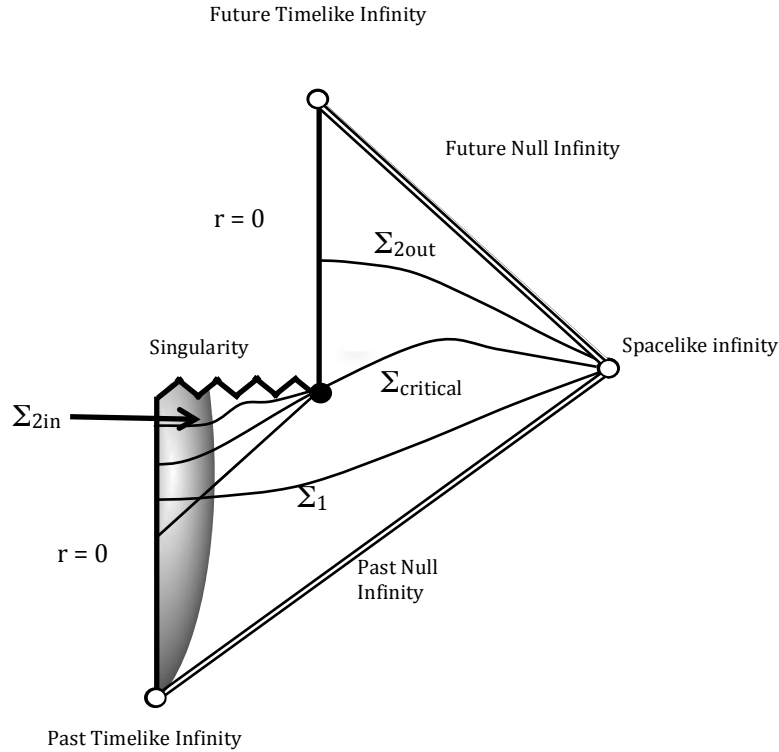

Figure 5: Foliating the evaporating black hole space-time

By using connected Cauchy surfaces until hitting the Evaporation Event and disconnected surfaces thereafter, the space-time can be foliated into Cauchy surfaces. Notice that the answer to the question "where is the information about the matter that fell into the black hole?" is exactly the same for $\Sigma_{2in} \cup \Sigma_{2out}$ as it was for $\Sigma_2$ in the non-evaporating case: the information is inside the event horizon, where the matter intersects the Cauchy surface. That's a natural place for the information to reside, and so it does. As soon as one asks a sensible question about where the information is one gets a sensible answer.

After the Evaporation Event, each Cauchy slice must have an "in" part and an "out" part. The set of "in"s foliate the interior of the event horizon above $\Sigma_{critical}$ and the set of "out"s foliate the region that is bounded by $\Sigma_{critical} \cup$ Future Null Infinity $\cup$ Timelike Infinity. Between the two of them, they lick the platter clean: the whole space-time gets foliated.

In specifying a foliation we must match each "in" piece with a particular "out" piece. The matching must preserve time order: if the "in" part of slice A precedes the "in" part of slice B then the "out" part of A must also precede the "out" part of B. And each set of slices must foliate the regions mentioned above. But aside from that, there is no constraint on how the "in"s get paired with the "outs". Foliations with different pairings are all equally legitimate.

We have no grounds from quantum theory to suspect that the dynamical transition from the physical state on $\Sigma_1$ to the physical state on $\Sigma_{2in} \cup \Sigma_{2out}$ is not deterministic, retrodictable and unitary[3]. Again, our problems have all evaporated. There must, nonetheless, be something unexpected about this resolution, since it has not been acknowledged as eliminating the paradox. At least one reason for the elusiveness of this resolution can be found in another of Geroch's theorems.

Not only does Geroch prove that any space-time that admits of a Cauchy surface admits of a foliation into Cauchy surfaces, he also proves that the topology of the Cauchy surfaces *cannot change*. But the topology of our Cauchy surfaces *does* change from connected to disconnected. What gives?

The solution to this puzzle can easily escape one's notice. Geroch's proof, like most proofs in this field, begins by characterizing a space-time as a "4-dimensional manifold with a smooth metric $g_{ab}$ of signature (+,-, -, -)". But our evaporating black hole space-time *is not a manifold*. The structure fails to be a manifold exactly because of the geometry at the Evaporation Event. Think of it this way. The "shrinking of the black hole" refers to the shrinking of the event horizon, and the event horizon is a topological 2-sphere. There is no topological discontinuity as the event horizon shrinks. But there is the unique critical space-time location where the event horizon disappears: the sphere shrinks down to a point. At exactly that point, the manifold structure fails. And any of Geroch's proofs that rely on the manifold structure obtaining everywhere have to be reconsidered in light of this.

One of those proofs, as we have seen, is that any space-time that admits of a Cauchy surface admits of a foliation into Cauchy surfaces. So we ought to check that this result still holds in our evaporating black hole space-time. But we have already done this: Figure 5 indicates how to foliate into Cauchy surfaces. On the other hand, we also know that the Cauchy surfaces have to change their topology once they are past the Evaporation Event (i.e. once the Evaporation Event lies in their past domain of dependence). That's the only way that the surface can intersect both the inextensible timelike curves that end up at Future Timelike or Null Infinity and those that end up at the singularity.

All of this is mathematically correct, but nonetheless it feels like something has gone wrong. To get to the bottom of the confusion we have to go beyond the mathematics and examine the intuitive concepts that are in play when people think and talk about the situation. But the observation that one of Geroch's theorems applies to this case and the other doesn't goes some way to explaining why this solution has remained elusive.

---

[3] There may be a problem for retrodiction from the Relativistic side. One can retrodict from the state on $\Sigma_{2in}$ what is in the past domain of dependence of $\Sigma_{2in}$ and similarly for $\Sigma_{2out}$.. But it is not evident how to determine how these two pieces fit together.

"But it no longer exists!"

Suppose you are watching the final stages of the evaporation of a black hole and the process is well described by the space-time structure of Figure 4. The optical distortion due to the black hole gets milder and milder, and covers a successively smaller area. It is like watching a balloon running out of air. The contraction accelerates and then finally, in a flash of light, it's gone. (The light is the light that had been on the event horizon, which passes through the Evaporation Event.)[4] After that, nothing. Space-time is smooth and empty. No mark that the black hole ever existed remains.

You ask: "Where did the information inside the black hole go?" and receive the answer I am advocating: the information is inside the event horizon. But how can that possibly be right: the event horizon has evaporated, it has disappeared, it no longer exists. How can what is non-existent contain anything at all?

Our problem here is not mathematical or physical but conceptual. In a relativistic setting, whether it involves black holes and evaporation or not, what precise physical meaning can be given to "X still exists" and "X no longer exists"? Once we get clear about that, all of these puzzles fade away.

"X no longer exists" is easy: in Relativity, "X no longer exists" means "the entirety of X lies in or on my past light cone". The Roman Empire no longer exists as I write this sentence because the entirety of events that constitute the Roman empire precedes me in time. The only objective, coordinate independent meaning that can be given to "precedes me in time" is "lies in or on my past light cone". This is perfectly clear and intuitive.

What about the phrase "X still exists"? Relativity makes a dent in our classical intuitions here. If the only three options we have are "X no longer exists", "X exists at present" and "X has yet to exist", and if the first means "The entirety of X is in or on my past light cone" and the last (by parallel reasoning) means "The entirety of X lies in or on my future light cone", then "X exists at present" must mean "Some part of X is space-like separated from me".  This does a little bit of violence to our usual intuitions, since it means that "what exists at present" has temporal thickness. On a planet far enough away, according to this usage, the entire lifetime of a creature from birth to death can "exist at present" for me.  In the classical regime, a "moment of time" has no temporal thickness, and what exists "at present" or "now" is what exists at that moment. But this is the only objective, coordinate independent meaning the phrase can have in Relativity.

If one wants to insist on having a "present time" or "present moment" with no temporal thickness in Relativity one can make use of a foliation into Cauchy surfaces. To say that something "exists at present" or "exists now" would be to say that part of it lies on the particular Cauchy surface on which the utterance also lies. The problem with this convention is that there are too many foliations by Cauchy surfaces that contain any given event, and one would need a method to pick out exactly which one is meant. No such general method exists.

---

[4] It is tempting to imagine a soft popping sound at that instant, but that is not supported by known physics.

(In Minkowski space-time there is one obvious method: let "now" for an object (which could be a person) at some event (which could be the utterance of "now") by the unique flat hyperplane orthogonal to the object's world-line at that event. If the object always moves inertially, then the associated set of "now"s foliate the space-time with flat hyperplanes. But if the object accelerates, or if the space-time is not flat, all bets are off concerning how these hyperplanes behave. They typically will not form a foliation.)

For our purposes, it makes no difference which of these ways of attaching meaning to "now" or "at present" one adopts. Either way we get the same result, namely that even though the *Evaporation Event* is in your past, it does not follow that the *interior of the black hole* is in your past. That can be read off the Penrose diagram. Pick any point in the future light cone of the Evaporation Event. No point in the interior of the event horizon lies in or on the past light cone of that event. The whole of the interior "still exists" according to the only sensible way we have to use those words. This is true even though the event horizon itself no longer exists: it is all in your past.

Shrinking black holes are not like shrinking balloons. If you could watch a balloon shrink down to a point and disappear, after the disappearance both the Disappearance Event *and the set of events that constitute the interior of the balloon throughout its existence* will lie in your past light cone. Both the balloon and its interior will no longer exist for you. One can easily fall into conceptual errors by assimilating the behavior of spherical objects and their interiors to the behavior of event horizons and their interiors. By being careful to separate these, we can answer puzzles. As quoted above, Hawking asks "So what would happen to all that information locked inside a black hole that evaporated away and disappeared completely?" [6]. Answer: it is still inside in the interior of the event horizon. Nothing happened to it.

It may also be useful to note an important difference in the logic of timelike relations and the logic of spacelike relations. Two events are timelike related if and only if they are connected by a timelike curve. So it is natural to establish timelike relatedness by seeking such a curve. The two pieces of $\Sigma_{2in} \cup \Sigma_{2out}$, being disconnected, somehow don't look to form a maximal locus of spacelike related events. But spacelikeness is not similar to timelikeness in this way: being connected by a spacelike curve is not criterial for being spacelike connected. Typically, every pair of events in a connected space-time can be connected by a spacelike curve. Spacelike relatedness is defined negatively: events are spacelike related if they are neither timelike nor null related. Using this criterion it is clear that the two parts of $\Sigma_{2in} \cup \Sigma_{2out}$ are indeed spacelike related.

## The External Observer

An entirely different set of confusions can arise from the use of phrases like "what the external observer can know" or "how things look to the external observer" or "can the external observer recover the information about what fell into the black hole?". The argument that Hawking radiation would be purely thermal, and hence carry no information (except the mass, charge and spin of the black hole) out to Null Infinity, is supposed to be the main source of the information loss paradox. But that only poses a problem if the information has to *get* to the "external observer". As we have seen, no one thought there was any information loss problem for the non-evaporating black hole, so there can't be any requirement that the information be accessible to an observer outside the event horizon.

In what sense should we expect that an external observer, i.e. an observer outside the event horizon, should be able to recover the initial state of our evaporating black hole system? Surely not in the sense that the observer can *do an experiment, even in principle, that would reveal the initial state*. That sort of information can never be recovered in quantum theory. Suppose, to take a simple example with no connection to black holes, an "outside observer" comes by an electron and wants to figure out what its *present* spin state is. No possible experiment she can do will reveal that information with certainty. If she wants to know how the electron was initially prepared, she is just out of luck.

We have understood "preservation of information" in terms of the dynamical connection between different global states of the universe, not in terms of information that can be *uncovered* or *"read"* by anyone. Demanding the latter is demanding too much. One could, of course, mean by "the external observer" just "the state on a slice that lies entirely outside the event horizon" without worrying about whether "the information" can be recovered. But as we have seen, there are no grounds to expect the state of such an exterior slice to carry all the information there is about the past.

Sometimes there is talk of what the external observer *sees*. This particular terminology is especially ambiguous. In one sense, it can be about what an observer literally sees, i.e. the light that gets to the observer. So, for example, one may be told that an outside observer never "sees" an object pass the event horizon of a black hole. In terms of what she literally sees this is correct. Since the light emitted at the event horizon stays on the event horizon it never reaches any observer that stays outside. But there is a completely different understanding of what "as seen by the outside observer" or "according to the outside observer" can mean.

It all starts with Special Relativity. It is often said that in Special Relativity an object appears shorter to an observer in relative motion with respect to it than to an observer at rest with respect to it. But this talk of how objects "appear" or "look" is not literally about how things appear to the naked eye. That can be determined by tracing the light rays that enter the eye. To such an observer, relatively moving objects literally appear not merely foreshortened but deformed and twisted in various ways. What is meant instead by "how things seem to the observer" is "how the physical state is described in terms of a coordinate system naturally associated with the observer". As we have already noted, in Special Relativity inertially moving objects are naturally associated with the Lorentz coordinate systems in which they are at coordinate rest. Clocks are "dilated" and rods "shortened" with respect to how they are described in various of these Lorentz coordinate systems. This sense of "seeming" has nothing to do either with how things literally look to the observer or with what the observer could determine by doing experiments.

Since there is no such canonical association of global co-ordinate systems with observers (whether inertial or not) in General Relativity, this locution has no obvious reading of this sort there. If one chooses a coordinate system whose t = constant surfaces form a foliation into Cauchy slices, one can ask how things "appear" in those coordinates. The "state at time $t_n$" means "the state on the Cauchy surface labeled by t = $t_n$". In this locution, if one asks where the "missing" information is at some time after the black hole evaporates, one is asking where the information is on a Cauchy slice that has the Evaporation Event in its past. And the answer, as we have seen, is "inside the event horizon".

Illustrative Example

We have identified errors that can be a source of confusion about evaporating black holes. An examination of the literature reveals instances of authors committing these errors and thereby falling into confusion.

Let's look at one of the foundational papers in the field: Susskind, Thorlacius and Uglam [8], which introduced the notion of black hole complementarity. In this case, we find exactly these problems in the introductory set-up of the paper. We are told "Consider a Penrose diagram for the formation and evaporation of a black hole, as in Fig. 1. Foliate the spacetime with a family of spacelike Cauchy surfaces, as shown" (p. 3744). The accompanying diagram is as follows (Figure 6):

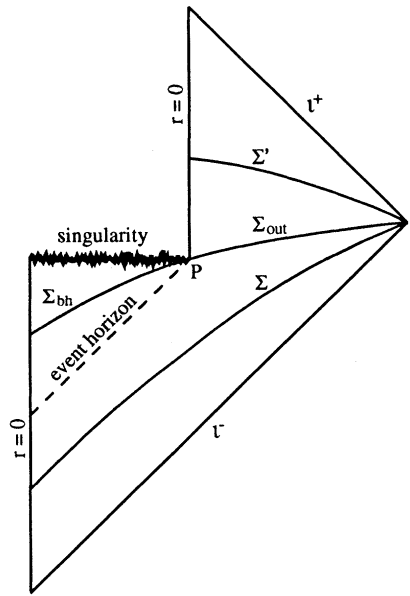

FIG. 1. Penrose diagram for black hole evolution.

Figure 6: From Thoralacios, Susskind and Uglam (1993)

Since Σ' is not Cauchy, the succeeding discussion is not pellucid.

Next, consider evolving the state further to some surface $\Sigma'$ in the future, as indicated in Fig. 1. The resulting state, $|\psi(\Sigma')\rangle$, represents the observable world long after the black hole has evaporated. According to Postulate 1, $|\psi(\Sigma')\rangle$ must be a pure state which is related to the original incoming state $|\psi(\Sigma)\rangle$ by a linear operator $S$, the $S$ matrix. By assumption, $|\psi(\Sigma')\rangle$ has evolved by the Schrodinger equation from some state $|\chi(\Sigma_{out})\rangle$ defined on $\Sigma_{out}$ which must then also be a pure state. This, in turn, implies that $|\psi(\Sigma_p)\rangle$ must be a product state:

$$|\psi(\Sigma_p)\rangle = |\Phi(\Sigma_{BH})\rangle \otimes |\chi(\Sigma_{out})\rangle$$

where $|\Phi(\Sigma_{BH})\rangle \in \mathcal{H}_{BH}$ and $|\chi(\Sigma_{out})\rangle \in \mathcal{H}_{out}$.

Already we are in deep trouble.

The conclusion that the state on the critical slice $\Sigma_p$ ($\Sigma_{BH} \cup \Sigma_{out}$) must be a product state is intolerable. What if the initial state contains a pair of entangled particles, one of which falls into the event horizon and the other of which does not? Surely such a thing Is physically possible. But then the state on $\Sigma_p$ will not be a product state.

The problem arises from the assumption that the state on $\Sigma'$ must be pure. And what, in turn, is this based on? Postulate 1 asserts that the transition from the initial state of infalling matter to the state of outgoing Hawking radiation must be implemented by a unitary S-matrix. It asserts in addition that "there exists a Hamiltonian which generates the evolution for finite times." Such a Hamiltonian should generate a unitary evolution from Cauchy surface to Cauchy surface. (It is somewhat puzzling that one would assume the existence of both an S-matrix and a Hamiltonian: surely, the evolution for finite times is all that is needed, since actual physical transition times are always finite.) This would yield the result that the purity of $|\psi(\Sigma)\rangle$ would entail the purity of $|\psi(\Sigma')\rangle$ *if* $\Sigma'$ were a Cauchy surface. And the original command to foliate the space-time into Cauchy surfaces misleadingly suggests that it is. Once we recognize that $\Sigma'$ is not a Cauchy surface there is no justification for expecting $|\psi(\Sigma')\rangle$ to be pure, and the rest of the puzzle cannot get off the ground.

The AMPS paper [9], which introduced the concept of firewalls, takes over the Susskind, Thorlacius and Uglam presentation of the problem verbatim, and so inherits the same weakness. Much of the seminal literature on the problem seems to have fallen prey to the same conceptual mistake.

## Remarks on the Evaporation Event

In the discussions so far we have described the space-time of the evaporating black hole as depicted in Figure 4. In particular, we have treated the point on the Penrose diagram labeled "Evaporation Event" as denoting an actual event in physical space-time. It is here that the manifold structure of the space-time fails, and with it the conditions for Geroch's theorems. Let's call this space-time structure *EE-in*.

One can see why the Evaporation Event must have an unusual topology in

Figure 7. This depicts the 1 + 1 dimensional space-time as a whole, not reducing the diagram by using the r = 0 axis as an axis of symmetry. The shaded region is the past light cone (with interior) of the Evaporation Event, with various null geodesics coming into and going out of the event as shown. Note how the past light cone of the Evaporation Event has a doubled structure while the future light cone is normal. Light whose trajectory is along the event horizon would presumably be combined with light coming in from outside the black hole as indicated by the various dotted and dashed lines. The two right-moving null geodesics would combine, as would the two left-moving ones.

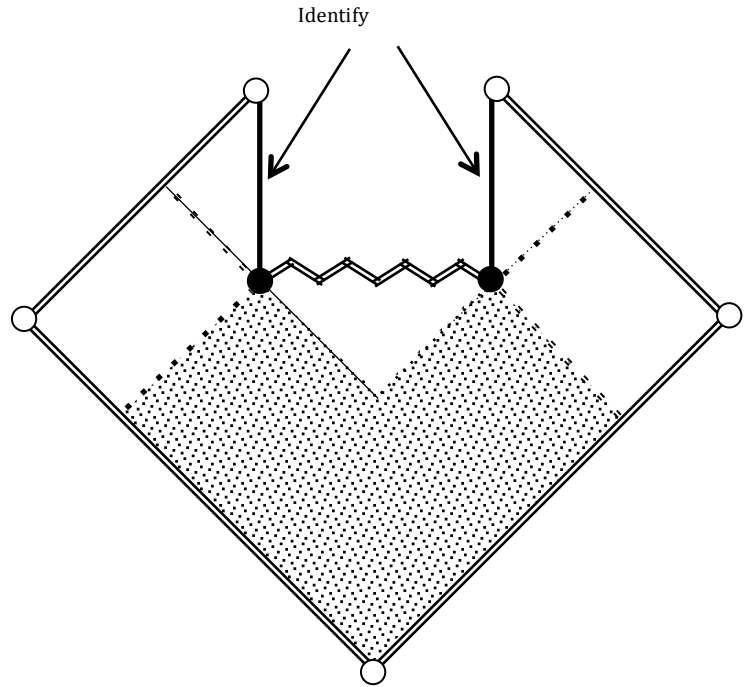

Figure 7: The Structure of EE-in

If we adopt this account of the space-time then we are faced with the task of specifying both the complete geometrical structure at the Evaporation Event and the physical laws that obtain there. This is sure to be a non-trivial task, although there are some clues to go on. As noted, it seem fairly obvious how the light cone structure must be formed at the Evaporation Event, which is all we make use of when settling claims about the Cauchy surfaces in the space-time. But even with those clues, there may be various plausible ways to specify the physics of the anomalous event.

There is an obvious way to avoid all of this extra work: just delete the Evaporation Event from the space-time altogether. This is the route taken by, e.g., Wald in [10], and is the natural reaction of any relativist accustomed to working only with 4-manifolds. But nothing comes for free. If we delete the Evaporation Event we get the situation depicted in Figure 8. Now light rays travelling along the event horizon cannot be continued beyond it, and there is a naked singularity where the Evaporation Event "used to be". Let's call this approach *EE-out*.

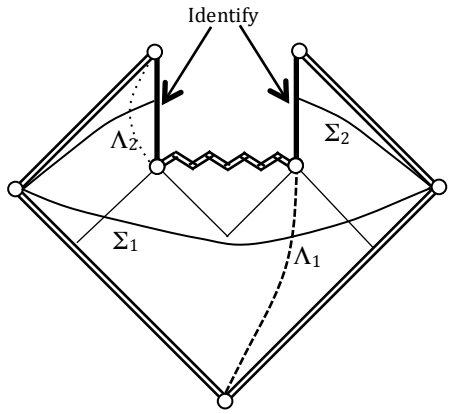

Figure 8: The Structure of EE-out

Because there is no physical Evaporation Event, both $\Lambda_1$ and $\Lambda_2$ are inextensible timelike curves. So it is easy to see that even $\Sigma_1$ is no longer a Cauchy surface: $\Lambda_2$ never intersects it. Indeed, in EE-out there are no Cauchy surfaces at all. The space-time is no longer globally hyperbolic and Geroch's theorems about Cauchy surfaces are moot. Intuitively, information can fall into the Evaporation Singularity along $\Lambda_1$ and be lost, and arbitrary states can emerge from the naked Evaporation Singularity along $\Lambda_2$. Because of the first feature, the state on $\Sigma_1$ cannot be retrodicted from the state on $\Sigma_2$, even if the latter is supplemented with information from within the event horizon. Because of the second feature, the state on $\Sigma_2$ cannot be predicted from the state of $\Sigma_1$.

*Prima facie*, then, one has no reason to expect determinism, or unitarity, or retrodictability, or conservation of information in EE-out. And it is not merely that people were mistaken when they identified $\Sigma_2$ as a Cauchy surface, they were mistaken when they identified $\Sigma_1$ as one.

Just as the proponent of EE-in owes us some new physics, viz. the physics of the Evaporation Event, the proponent of EE-out owes us the new physics of a bare singularity. The benefits and burdens of the two approaches are different, and it is not clear which is the more promising avenue of research.

When Wald discusses EE-out in [10] he suggests a way to have his cake and eat it too. Even though it seems as if the dynamics of EE-out cannot be deterministic, Wald argues that it can be. Using an analogy, he considers solving the wave equation for a scalar field on a Minkowski space-time from which a single point has been deleted. The modified space-time is no longer globally hyperbolic and there are no Cauchy surfaces. But, he claims, continuity considerations can force a unique solution to an "initial condition" problem nonetheless.

It is an open question whether Minkowski space-time with and without a single point is really a good analogy for the difference between EE-in and EE-out. In particular, in the Minkowski case the "missing point" can be supplied in only one reasonable way to yield a manifold. The problem with EE-out, in contrast, is that the addition of a single point cannot yield a manifold. So Wald's claim needs more investigation.

There is another line of thought that seems plausible with regard to both EE-

in and EE-out. The thought is that the presence or absence of the Evaporation Event can't make *that much* difference with respect to information transmission from before the Event to after. It is, after all, only a single *point* that is either missing (in EE-out) or non-manifold (in EE-in). And a point can't absorb or emit very much information: it's just too small. So not much information can be absorbed by the singularity or emitted by the naked singularity in EE-out. Determinism, unitarity and information conservation ought to *nearly* obtain. And in any case, the account of information about *what fell in to form the black hole in the first place* seem to be completely untouched by these issues. That information is inside the black hole, and registers on a partial Cauchy slice like $\Sigma_{2in}$, no matter what happens at the Evaporation Event.

This argument needs to be fleshed out. Certainly, with respect to pure-to-mixed transitions it is not at all true. Start with a large collection of particles in a maximally entangled state, and then lose just a single particle to the Evaporation Singularity. The remaining state is as mixed as it could be, and a tremendous amount of information has been "lost" because there are so many distinct maximally entangled states one could have started with.

A more generalized idea that there is a direct relation between the size of a region of space-time and the amount of information it can contain is pursued in the next section.

## Spatial Volume Bounds on Information

There is one other avenue of argument to the existence of an "information loss paradox" when a black hole evaporates. I have argued that there is no problem of information loss because full information is retained on all Cauchy slices, which is all we ever had any reason to expect. In conversation, I have heard this objection to the view: as the black hole shrinks its interior becomes ever smaller, there is just not enough *room* inside to hold the requite information. It must either leak out somehow or be destroyed. If there is no loss it must leak out, so we are back to asking how it escapes the event horizon.

In order to make clear sense of this objection at least four tasks must be performed. First, one must somehow *quantify* the information, so that it makes sense to discuss how much of it there is in a system. Second, one must *locate* the information, so one can determine how much is contained in some volume of space or space-time. Third one must *define the volume of the interior of the event horizon*, so it makes sense to say how much room there is for information to fill up. And finally one must *justify the bound* put on the amount of information that can be contained in a volume. If this sort of objection can be made valid, all of our observations about Cauchy surfaces will be for naught. Having said that the information must be inside the event horizon, we are driven to the conclusion that it can't be.

Each of these four tasks is problematic. The usual way to quantify information is via Shannon's theory. But that theory requires a division of the physical situation into a sender, a receiver, and an informational channel. The sender has to be characterized by a probability distribution over possible transmission states. The channel has to be characterized by parameters of proper operation. None of this additional conceptualization falls immediately out of physical principles. So the quantification problem is not trivial.

The location problem is even worse. It is easy to describe cases where a system unproblematically "holds information" but there is no obvious physical location where it is held. This happens even classically, and is exacerbated by quantum theory.

A classical example. Alice wants to send a 1-bit message to Bob, but is worried about it being intercepted and decoded. She has two couriers at her disposal. There is a danger that one or the other of the couriers will be caught, but no danger both will be. Alice wants to be sure that neither courier carries any information about her message to Bob, but both together carry the single bit.

Alice and Bob agree to this scheme. If Alice's message is "yes" she will send the couriers with matching colored slips of paper, either both red or both green. If the answer is "no" the colors will be different. Alice will flip a fair coin to decide whether Courier 1 gets a red paper or a green paper. In Shannon's terms neither courier's piece of paper carries any information about Alice's message, but both together contain perfect information. So just where was the information as the couriers made their way from Alice to Bob? It is not located with either courier alone. If we ask how much information is contained in a particular courier's present location it is not clear how to answer.

The problem gets even worse with quantum theory due to entanglement. Now Alice can implement the following scheme: if her answer is "yes" she prepares a pair of electrons in the singlet state, and if it is "no" she prepares the m = 0 triplet state. She gives one electron to each courier and sends them on their way. Now the capture of one courier can be of no possible use: no matter which message is being sent the reduced density matrix of each electron will be exactly the same. Experiments carried out either electron alone can therefore in principle not reveal any information at all about the message. They have the same statistical predictions irrespective of the message. But again the single bit of information gets to Bob. It starts with Alice and ends with Bob, but has no definite location in space in the interim. So the location problem is non-trivial.

What about the volume of the interior of the event horizon? Suppose the event horizon has "shrunk" to an area of $4\pi$ cm$^2$. If the black hole were a Euclidean sphere with this surface area it would have an interior volume of $\frac{4}{3}\pi$ cm$^3$. But we obviously cannot ascribe that volume to the interior of the event horizon. First, the space-time is highly curved, so the use of the Euclidean relation is unjustifiable. Second, in order to have a spatial object whose volume is being measured we have to fix a spacelike hypersurface bounded by the event horizon. Different such surfaces will yield different volumes, as measured by the Riemannian metric induced on the hypersurface.

If there is a spacelike hypersurface of maximal volume, it would be safe to use that one. The relevant volume may, though, be quite difficult to calculate since the hypersurface will pass through the infalling matter, so the geometry is likely to be quite complicated.

Suppose that the first three problems have been solved, so we can locate the information inside the event horizon, quantify it, and define the volume of the interior. We are left with the hardest problem of all: justifying a bound that the volume puts on the amount of information that can be contained in it. For it is not merely that General Relativity itself implies no such bound, but rather than the fundamental principles we have to hand *are incompatible* with any such bound.

In both classical General Relativity and quantum field theory (without collapse of the quantum state) information is always conserved in any Cauchy-to-Cauchy transition. That's because the information on each Cauchy slice is maximal information: from it the entire physical state over all time follows. So we have the principle that *all Cauchy slices in these two theories contain the same amount of information*.

To make things simple, let's consider a case set in Minkowski space-time. Constructing a parallel argument in a General Relativistic space-time is trivial. In a given set of Lorentz coordinates, the flat hyperplane t = 0 is a Cauchy surface. Divide the hyperplane into equal-volume chunks of, e. g. ,1 cm in a 1 +1 spacetime, 1 cm² in a 2 + 1 space-time, 1 cm³ in 3 + 1, etc. Now put some physical state, which contains a particular amount of information, on this Cauchy slice. If all of our other problems have been overcome, we should be able to locate every bit of information inside some particular chunk, and count the number of bits in each chunk. Now suppose there is a principle that bounds the amount of information that each chunk can contain. That supposition leads to a contradiction with our other principles.

It is easiest to see this in the 1 +1 case, but the argument easily generalizes to higher dimensions.  Figure 9 depicts a 1 + 1 dimensional space-time with three different Cauchy surfaces. The surfaces overlap outside the 1 cm interval marked, and in that interval they diverge from the flat surface by varying amounts. The length of the surfaces in that interval get successively smaller the more they diverge, with the length limiting to 0 as the surface approaches the null surfaces indicated.

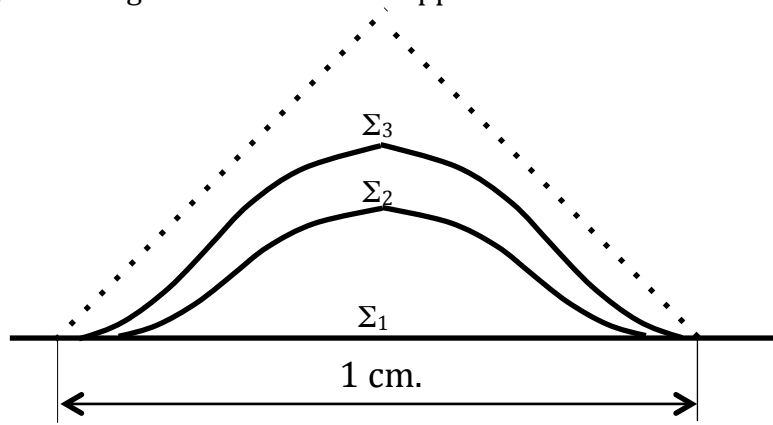

Figure 9: Three Cauchy surfaces

According to our principles, the three Cauchy surfaces contain the same amount of information, which information can be localized on the Cauchy surface. Since the surfaces all overlap outside the interval they contain the same amount of information there. It follows that they contain the same amount of information also inside the interval. But the available volume on the surfaces for storing information inside the interval is different and can be made as small a positive amount as you like. So either the principle that all Cauchy surfaces contain the same information, or the principle that the information can be located of the surface or the principle that there is a bound on how much information can be contained in a volume must be false.

Talk of informational bounds for volumes tends to go hand-in-hand with the idea that space-time itself will be quantized in quantum gravity. All such theories are highly speculative, and it should also be noted that not all quantities become discretized in quantum theory. The energy of a free particle in quantum mechanics, for example, is not discretized. So this form of argument for an information loss problem depends both on denying some of the properties of existing theories and imposing a particular condition on the not-yet-discovered theory of quantum gravity. This logical situation undercuts any pretension that the "information loss paradox" holds clues of how to proceed in quantizing gravity. As we have seen, there is no problem at all if one retains the properties of the theories we have in hand. And to get a problem one not only has to abrogate those properties but also to import assumptions about the very theory we are seeking. These are thin reeds on which to base the heuristics for quantum gravity.

## AdS/CFT and Superscattering

There are a couple of loose ends to be briefly addressed. One is the bearing of the AdS/CFT conjecture on the argument of this paper. That conjecture has been widely regarded as supporting the conclusion that all the information about what initially fell into the black hole must, somehow, escape from the black hole in the course of evaporation. For there are no singularities on the boundary of the Anti-deSitter space-time, so the dynamics of the conformal field theory there must be unitary and deterministic and retrodictable. Since there is supposed to be a 1-to-1 correspondence between states on the boundary and states of the bulk, the bulk dynamics must also display those features.

Even if we accept the conjecture and its extension to physically realistic cases, though, the conclusion does not follow. As we have seen, we can have unitary, deterministic, retrodicable transitions from the state on one Cauchy surface to the state on any other without the information ever "escaping" from the interior of the event horizon. Furthermore, in the absence of any understanding of how the space-time geometry of the bulk is "encoded" in the degrees of freedom at the surface it is impossible to use these considerations to determine the geometry of the evaporating black hole space-time in the bulk. So with regard to the question of whether the "information" ever "escapes" from the black hole, the AdS/CFT conjecture is quite useless.

The other loose end is Hawking's [11] suggestion that the moral of the information-loss argument is that we should be using superscattering operators—which operate on density operators rather than only pure states and allow for the transition from pure to (improper) mixed states—rather than scattering operators in calculating the transition from the state on $\Sigma_1$ to the state on $\Sigma_2$. We have argued that the transition from the state on $\Sigma_1$ to the state on $\Sigma_2$ (i.e. $\Sigma_{2out}$) can indeed be pure-to-mixed. There is a pure-to-pure transition from $\Sigma_1$ to $\Sigma_{2out} \cup \Sigma_{2in}$ (governed by the fundamental Hamiltonian) followed by tracing out the $\Sigma_{2in}$ part. The question is which superscattering operator would implement that transition.

In pursuing this question, Hawking postulates some physical principles that the transition encoded in the operator must validate. Of course, he demands that the transition from a state on $\Sigma_1$ to a state on $\Sigma_2$ display these properties. One is unitarity, so that the transition amplitudes can be interpreted as probabilities. Another is the global conservation of energy, so the total energy ascribed to the initial state on $\Sigma_1$ must be the same as the total energy ascribed to the final state on $\Sigma_2$. The first condition is surely required: given any initial state on $\Sigma_1$, the sum of the probabilities of all possible final states on $\Sigma_2$ must be unity. But given that $\Sigma_2$ is not Cauchy, the justification for the second condition is not so evident. In the transition from the state on a Cauchy surface to the state on a non-Cauchy surface there is no obvious reason why energy should not be lost. Furthermore, the whole question of the meaning and validation of any principle of global conservation of energy in this situation is somewhat vexed. But that is the topic for another time.

## Peroration

For over forty years, the information loss paradox has held a prominent place in discussions of the prospects for a quantum theory of gravity. After so much time going unsolved, proposals have become ever more outlandish. Black hole complementarity, firewalls, and EPR = ER are examples of proposals based on principles that are conjectural and shaky at best and fanciful or incoherent at worst. This remarkable situation has a simple if unexpected explanation: there is no problem that needs to be solved. Once one is careful and precise about the principles at issue, such as retrodictability and unitarity and the meaning of information conservation, one finds that there is no reason to think they were ever violated in the first place. Even if black holes shrink and eventually evaporate the fundamental principles can all obtain. The main thing to be learned is that the evaporation requires a very localized failure of the manifold structure of space-time at the Evaporation Event. This in turn renders a key theorem about Cauchy surfaces inoperative, allowing them to change their topology. This unexpected behavior, coupled with conceptual confusions about temporal terminology, seems to have obscured the straightforward response to the supposed paradox.

When we run over the literature on the paradox, persuaded of these principles, what havoc shall we make? If we take in our hand any article, from *Physical Review D* or *General Relativity and Gravitation*, for instance, let us ask: *Does it refer only to states on Cauchy surfaces?* No. *Has it explained why information or purity should be preserved in a Cauchy-to-non-Cauchy transition?* No. Commit it then to the flames; for it can contain nothing but sophistry and illusion.

Acknowledgements

Thanks to discussions and correspondence with David Albert, Sean Carroll, John Earman, Robert Geroch, Shelly Goldstein, Ned Hall, Sabine Hossenfelder, Vishnya Maudlin, Bert Sweet and Robert Wald.

Bibliography

[1] S. W. Hawking, "Particle creation by black holes", Commun. Math. Phys. **43**, 199—220 (1975)

[2] P. S. Laplace, *A Philosophical Essay on Probabilities*, translated into English from the original French 6th ed. by Truscott,F.W. and Emory,F.L., Dover Publications (New York, 1951)

[3] B. Kožnjak, "Who let the demon out? Laplace and Boscovich on determinism", Studies in History and Philosophy of Science **51** 42-52 (2015)

[4] R. Wald, *Quantum Field Theory in Curved Spacetime and Black Hole Thermodynamics,* Chicago University Press (Chicago 1994)

[5] N. Huggett, ed. *Space from Zeno to Einstein*, MIT Press (Cambridge, MA 1999)

[6] S. W. Hawking, "Information Loss in Black Holes", Phys. Rev. D **72**:084013 (2005)

[7] R. Geroch, "Domain of Dependence", J. of Math. Phys. **11** 2, 437-448 (1970)

[8] L. Susskind, L. Thorlacius, and J. Uglam, "The stretched horizon and black hole complementarity", Phys. Rev. **D48** 8, 3743-61 (1994)

[9] A. Almheiri, D. Marolf, J. Polchinski and J. Sully, "Black holes: complementarity or firewalls?", J. High Energ. Phys. **62** (2013)

[10] R. Wald, "Black Holes, Singularities and Predictabiity" in *Quantum Theory of Gravity*, edited by S. Christensen, Adam Hilger (Bristol, 1984) 160-8

[11] S. Hawking, "Breakdown of probability in gravitational collapse", Phys. Rev. D **14** 10, 2460-73 (1976)